\documentclass[prl,twocolumn,showpacs,preprintnumbers,amsmath,amssymb]{revtex4}
\usepackage{graphicx,graphics,wrapfig,rotating}         
\usepackage{epsfig}
\usepackage{dcolumn}                                    
\usepackage{bm,fancybox}
\usepackage{amsmath,amssymb}                          
\usepackage{times,euscript,oldgerm}              %
\usepackage[english]{babel}    
\usepackage{multirow}                  %
\usepackage{psfrag}
\usepackage{color}
\usepackage{longtable}

\newcommand{\Yb}{\ensuremath{^{171}\mathrm{Yb}^+~}}

\newcommand{\avg}[1]{\ensuremath{\left\langle#1\right\rangle}}
\newcommand{\ket}[1]{\ensuremath{\left|#1\right\rangle}}

\begin{document}

\title{State-independent experimental tests of quantum contextuality in a three dimensional system}

\author{Xiang Zhang$^1$, Mark Um$^1$, Junhua Zhang$^1$, Shuoming An$^1$, Ye Wang$^1$, Dong-ling Deng$^{1,2}$, Chao Shen$^{1,2}$, Luming Duan$^{1,2}$, Kihwan Kim$^1$}
\email{kimkihwan@gmail.com}

\affiliation{
 $^1$ Center for Quantum Information, Institute for Interdisciplinary Information Sciences, Tsinghua University, Beijing 100084, P. R. China
 $^2$ Michigan Center for Theoretical Physics and Department of Physics, University of Michigan, Ann Arbor, MI  48109, USA
}

\date{\today}
\begin{abstract}
We experimentally observed state-independent violations of Kochen-Specker inequalities for the simplest indivisible quantum system manifesting quantum contextuality, a three-level (qutrit) system. We performed the experiment with a single trapped \Yb ion, by mapping three ground states of the \Yb ion to a qutrit system and carrying out quantum operatations by applying microwaves resonant to the qutrit transition frequencies. Our results are free from the detection loophole and cannot be explained by the non-contextual hidden variable models.
\end{abstract}

\maketitle
It is a long-standing problem whether the nature of physical system would be completely described by quantum mechanics. In classical views the measurement outcomes on physical properties are non-contextual, $\it{i.e.}$, predetermined independently of their own and other simultaneous compatible measurements, while quantum mechanics is contextual. Kochen, Specker and Bell proved that quantum mechanics and any non-contextual classical theory are in conflict \cite{KS67,Bell66}, deeply rooted in the essence of quantum mechanics regardless of states of the system. The original logical proof has been formulated to experimentally testable inequalities, called Kochen-Specker (KS) inequalities. The Bell's inequalities that hold for classical theories with local hidden variables can be considered as a special type of KS inequalities, where the contextuality is presented by non-locality. While Bell's inequalities can be violated by special entangled states in space-like separation, the violation of KS inequalities could be observed by any quantum state in a system with dimension $d \geq 3$. For cases where $d \geq 4$, KS inequalities were proposed \cite{Mermin90,Peres91,Cabello08} and demonstrated in both a state-dependent \cite{Zukowski00,Guo03,Hasegawa09} and a state-independent manner \cite{Roos09,Cabello09,Laflamme10}. For the smallest case $d=3$, the state-dependent inequality was developed \cite{Shumovsky08} and tested with a photon system \cite{Zeilinger11}. Lately, the state-independent inequality was found \cite{Oh12,Larsson12} and the violations were informed with also a photon system \cite{Luming12}. However, these experimental demonstrations are open to the detection loophole. Here we report the experimental results of a single three-level atomic ion that are in conflict with non-contextual classical theories. The experimental violations of KS inequalities are observed independently of entanglement or superposition of the states and confirm the quantum contextuality at the most simple and fundamental level without the fair sampling assumption. 

In noncontextual classical models, values of an observable are determined by only a hidden variable independently of the measurement context, $\it{i.e.}$, set of mutually compatible observables measured in a single experimental setting. The observables are compatible if their results are not dependent on the order of measurements. Kochen and Specker showed that any quantum state in larger than 2 dimensions would reveal conflict with the non-contextual theories. The demonstration of the conflict in a simplest three dimensional quantum system, a qutrit, has fundemental importance since it would naturally imply the contraction in higher dimensions. However, the original proof of KS theorem for a qutrit involves too many configurations, 117 observables, to be verified in experiment \cite{KS67}. Differently from the KS proof, Yu and Oh realized a KS inequality of a qutrit with 13 observables and 24 correlations of them \cite{Oh12}. It has been proved that 13 settings are minimum for  KS tests of $d=3$ system \cite{Cabello11} and the bound have improved to be tight, which is favorable in experimental realization \cite{Larsson12}. 

\begin{figure}
\includegraphics[width=1\linewidth]{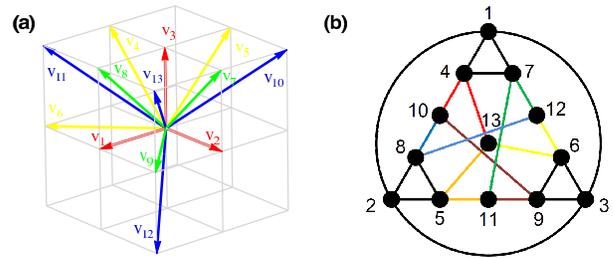}
\caption{ \label{fig1:13} Observables and compatiblity relations for the state-independent inequality of a qutrit. (a) The 13 observables are represented as vectors in a 3-dimensional space. $\ket{v_{i}}$s are specified as $v_{1}= \left(1,0,0\right)$, $v_{2}= (0,1,0)$, $v_{3}= (0,0,1)$, $v_{4}= (0,1,-1)$, $v_{5}= (1,0,-1)$, $v_{6}= (1,-1,0)$, $v_{7}= (0,1,1)$, $v_{8}= (1,0,1)$, $v_{9}= (1,1,0)$, $v_{10}= (-1,1,1)$, $v_{11}= (1,-1,1)$, $v_{12}= (1,1,-1)$, and $v_{13}= (1,1,1)$. (b) The compatibility graph between 13 observables. The nodes represent the 13 vectors $v_{i}$ and the edges show the orthogonality or compatibility relations. The experimental realization and joint measurements of the orthogonal observables are described in Table \ref{Tab1:Meassequence}. }
\end{figure}
    
The inequalities with 13 observables are represented in the Hilbert space of a qutrit with basis $\left\{ \ket{1}, \ket{2}, \ket{3} \right\}$. The form of the observable is $A_{i} = I- 2 V_{i}$, where $V_{i}$ is the normalized projection operator on a vector $\ket{v_{i}} = a \ket{1} + b \ket{2} + c \ket{3}$. The 13 vectors required for the inequality are displayed in 3$d$ Hilbert space in Fig. \ref{fig1:13}(a), where the values $(a,b,c)$ of a vector, $\ket{v_{i}}$ are assigned as real numbers, which are enough to cover all possible projections in a qutrit space. In Fig. \ref{fig1:13}(b) a edge $\left(i,j\right) \in E$ shows the orthogonal relation of nodes $i,j \in V$, where $A_{i}$ and $A_{j}$ are compatible observables. There are 24 edges requiring simultaneous measurements. The tight ineqaulity is written as \cite{Larsson12},   
\begin{eqnarray}
\avg{\chi_{13}} &=& \sum_{i \in V}  \mu_{i} \avg{A_{i}} - \sum_{\left(i,j \right) \in E} \mu_{ij} \avg{ A_{i} A_{j} }\nonumber \\ &-& \sum_{\left(i,j,k \right) \in C} \mu_{ijk} \avg{ A_{i} A_{j} A_{k}} \leq 25, 
\label{eq1:OhYu}  
\end{eqnarray}
where $\avg{\ldots} $ denotes the average of measurement outcomes, $\mu_{i}=1~ \left(i=1,2,\cdots,9\right)$, $2 ~\left(i=10, \cdots, 13\right)$, and $\mu_{ij}=1$, when $\left(i, j\right)$ are in the triangles $\left\{(1, 4, 7), (2, 5, 8), (3, 6, 9)\right\} $, $2$ otherwise. Here $C$ is the set of four triangles in Fig. \ref{fig1:13}(a) and $\mu_{ijk}=3$, when $\left(i, j, k\right) \in \left\{(1, 4, 7), (2, 5, 8), (3, 6, 9)\right\} $ and $0$ otherwise. The inequality is easily derived from an equation of the same form with 13 arbitrary variables that have $\pm 1$. In quantum mechanics, $\chi_{13} = \left( 25 + \frac{8}{3} \right) I$, which clearly violates the inequality (\ref{eq1:OhYu}) with any input states in the 3$d$ Hilbert space.

For the hidden variable models that preserve algebraic structures of compatible observables, the inequality would be simplified to \cite{Oh12},
\begin{equation}
\avg{ \chi_{4}} = \sum_{i =10}^{13} \avg{ V_{i}} \leq 1. 
\label{eq3:OhYuSimple}  
\end{equation}
The inequality is valid, assuming the $\emph{product rule}$ and the $\emph{sum rule}$ still hold. The $\emph{product rule}$ means the product of compatible observables are zero and the $\emph{sum rule}$ means the sum of all compatible observables are one. The $\chi_{4}= \left( 1 + \frac{1}{3} \right) I$ for the quantum mechanics, which breaks the inequality (\ref{eq3:OhYuSimple}) regardless of initial states.   

\begin{figure}
\includegraphics[width=1\linewidth]{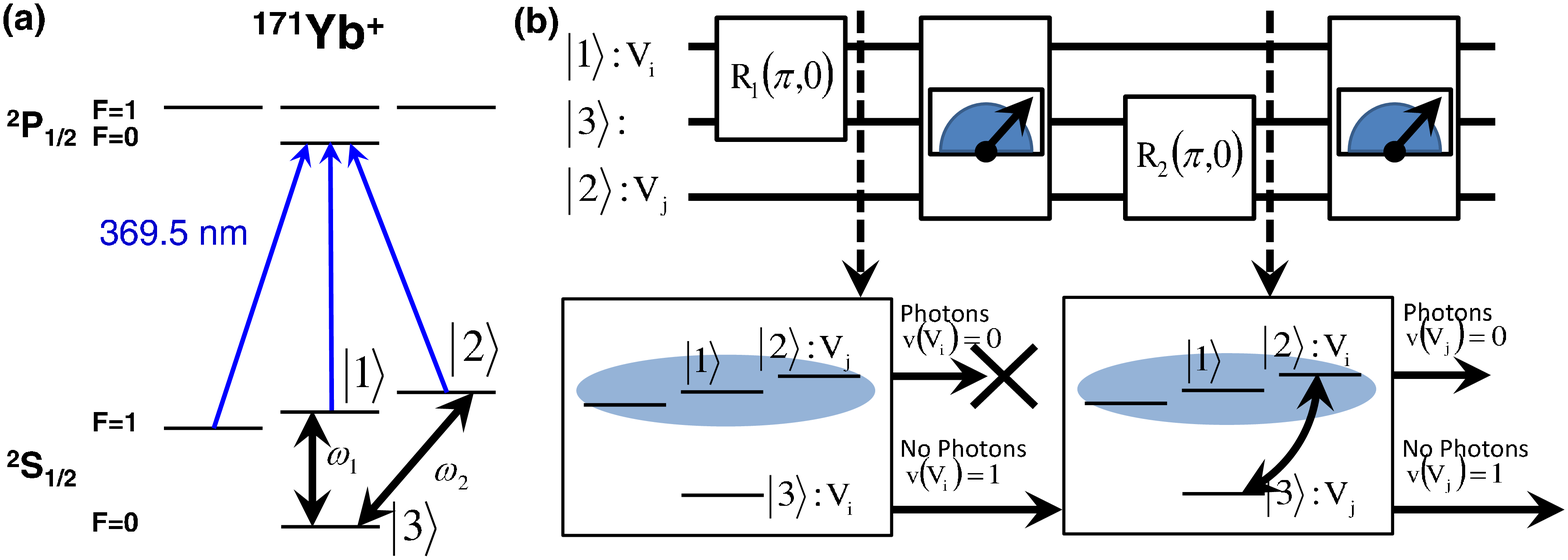}
\caption{ \label{fig2:Yb} The \Yb ion system and the measurement scheme. (a) The energy diagram of \Yb. Qutrit states $\ket{1}$, $\ket{2}$, and $\ket{3}$ are mapped onto $\ket{F=1, m_F=0}$, $\ket{F=1, m_F=1}$, and $\ket{F=0, m_F=0}$ in the $S_{1/2}$ ground state manifold, respectively. The transition frequencies are: $\omega_1= (2 \pi) 12642.8213$ MHz and $\omega_2 = \omega_1 + (2 \pi) 7.6372$ MHz. The quantum operations are performed by the resonant microwaves. The detection laser beams are shown as blue lines. Here the quantum state projected to $\ket{1}$ or $\ket{2}$ generates fluoresence, while the state collapsed to $\ket{3}$ does not generate photons. Therefore, we assign zero on the observable related to state $\ket{3}$ when we detect photons and one when no photons are detected. (b) The sequential measurement scheme to construct the correlation terms $\avg{V_{i}V_{j}}$. The three lines of the upper part stand for the three levels of \Yb ion. First, we rotate the measurement axis to map the observables $V_{i}$ and $V_{j}$, for example, on the state $\ket{1}$ and $\ket{2}$, respectively. We transfer the $V_{i}$ on the state $\ket{3}$ and apply the standard flueorescent detection scheme. If we detect photons, the state should not be $\ket{3}$ and we assign zero on the observable $V_{i}$. Therefore the correlation $V_{i}V_{j}$ should be zero and no further experiments are necessary. If we detect no photons, we assign one on the $V_{i}$. To measure the value of $V_{j}$, we apply the swapping microwave pulse that converts $V_{j}$ to $\ket{3}$ and the detection laser beam. If we observe photons, $V_{j}=0$ and if no photons, $V_{j} = 1.$ By repeating the same experimental sequences and counting the number of experiments that detect no photons at the second measurement, we can obtain the average of the correlations $V_{i} V_{j}$. We perform these sequencial detection schemes on the all the measurement settings listed in Table \ref{Tab1:Meassequence} for each initial state.}
\end{figure}

We performed the tests of the KS inequalities (\ref{eq1:OhYu}) and (\ref{eq3:OhYuSimple}) with a single trapped \Yb ion in a four-rods radio frequency trap similar to the one in Ref. [\cite{Olmschenk07}]. A trapped ion system would be an exemplary candidate for the test of the quantum contextuality, since the system has already demonstrated near-perfect state initialization, state detection, and quantum operations on a qubit system \cite{DidiRMP,Haffner08, ActonThesis}. A pair of qubits in a trapped ion system were already used to study the quantum contextuality of 4$d$ Hilbert space \cite{Roos09}. We map the three internal levels of the \Yb ion in the $S_{1/2}$ ground state manifold to the qutrit states, which are represented by: $\ket{1} = \ket{F=1, m_F=0}$, $\ket{2} = \ket{F=1, m_F=1}$, and $\ket{3} = \ket{F=0, m_F=0}$ shown in Fig \ref{fig2:Yb}(a). 

The initialization of the qutrit system can be realized by the same standard optical pumping as on a qubit. The operation on a single qutrit would be performed through the application of either resonant microwaves or lasers, depending on transitions. The quantum operations on internal levels of an atomic ion have been extremely well developed and demonstrated to a near-perfect level \cite{Wineland11}. However, it is challenging to simultaneously discriminate between three states of a qutrit with decent detection fidelity in trapped ion systems. The standard detection method based on qubit-state-dependent fluorescence enables us to differentiate between one state versus the other two states of a qutrit. We observe on average 10 photons for the $\ket{1}$ or the $\ket{2}$ state and detect no photon for the $\ket{3}$ state. And by exchanging populations between $\ket{1}$ and $\ket{3}$ (or $\ket{2}$ and $\ket{3}$), we could discriminate $\ket{2}$ and $\ket{3}$ versus $\ket{1}$ (or $\ket{1}$ and $\ket{3}$ versus $\ket{2}$. However it is not possible to simultaneously distinguish the three states at our current experimental setting. In summary, we assign the value 1 on $\ket{3}$ state when no photons are detected and put the value 0 when photons are detected.

\begin{table}
\renewcommand{\arraystretch}{0.5}
\centering
\scalebox{1}{
\begin{tabular}{ l || l | l | l || l }
  \hline                        
  & $\left|1 \right\rangle$ & $\left|2 \right\rangle$ & $\left|3 \right\rangle$ & Rotations Sequences ($1^{st}$ $2^{nd}$ $3^{rd}$) \\ \hline
  $M_1$ & $V_1$ & $V_2$ & $V_3$ & no rotation \\
  $M_2$ & $V_5$ & $V_2$ & $V_8$ & $R_1\left(\pi/2,\pi\right)$ \\ 
  $M_3$ & $V_1$ & $V_4$ & $V_7$ & $R_2\left(\pi/2, 0\right)$ \\
  $M_4$ & $V_9$ & $V_3$ & $V_6$ & $R_2\left(\pi, \pi\right) R_1\left(\pi/2,\pi\right) $\\  
  $M_5$ & & $V_4$ & $V_{10}$ & $R_2\left(\pi/2, 0\right) R_1\left(\alpha, 0\right) $ \\     
  $M_6$ & & $V_4$ & $V_{13}$ & $R_2\left(\pi/2, 0\right) R_1\left(\alpha, \pi\right) $\\ 
  $M_7$ & $V_5$ & & $V_{11}$ & $R_1\left(\pi/2, \pi\right) R_2\left(\alpha, \pi\right) $ \\   
  $M_8$ & $V_5$ & & $V_{13}$ & $R_1\left(\pi/2, \pi\right) R_2\left(\alpha, 0\right) $ \\ 
  $M_9$ & $V_6$ & & $V_{12}$ & $R_2\left(\pi, 0\right) R_1\left(\pi/2, \pi\right) R_2\left(\alpha, 0\right)  $ \\ 
  $M_{10}$ & $V_6$ & $V_{13}$ & & $R_2\left(\pi, \pi\right) R_1\left(\pi/2, 0\right) R_2\left(\pi-\alpha, 0\right)  $ \\ 
  $M_{11}$ & & $V_7$ & $V_{11}$ & $R_2\left(\pi/2, \pi\right) R_1\left(\alpha, \pi\right) $ \\ 
  $M_{12}$ & $V_{12}$ & $V_7$ &  & $R_2\left(\pi/2, \pi\right) R_1\left(\pi-\alpha, \pi\right) $ \\ 
  $M_{13}$ & $V_8$ & & $V_{10}$ & $R_1\left(\pi/2, 0\right) R_2\left(\alpha, 0\right) $ \\ 
  $M_{14}$ & $V_8$ & $V_{12}$ & & $R_1\left(\pi/2, 0\right) R_2\left(\pi-\alpha, 0\right) $ \\ 
  $M_{15}$ & $V_{10}$ & $V_{9}$ & & $R_1\left(\pi, 0\right) R_2\left(\pi/2, 0\right) R_1\left(\pi-\alpha, 0\right)  $ \\ 
  $M_{16}$ & & $V_9$ & $V_{11}$ & $R_1\left(\pi, \pi\right) R_2\left(\pi/2, \pi\right) R_1\left(\alpha, 0\right)  $ \\ 
\hline
\multicolumn{5}{c}{$R_1\! \left(\theta_1\!,\phi_1\!\right)\! =\! \left(\begin{array}{ccc}
\!\text{\ensuremath{\cos}}\frac{\theta_1}{2} \!& 0 & e^{i\phi_1}\text{\ensuremath{\sin}}\frac{\theta_1}{2}\\
\!0\! & 1 & 0\\
\!-e^{-i\phi_1}\text{\ensuremath{\sin}}\frac{\theta_1}{2}\! & 0 & \text{\ensuremath{\cos}}\frac{\theta_1}{2}
\end{array}\right)$} \\ 
\multicolumn{5}{c}{$R_2\! \left(\theta_2\!, \phi_2\! \right)\! =\! \left(\begin{array}{ccc}
\!1 & 0 & 0\\
0 & \text{\ensuremath{\cos}}\frac{\theta_2}{2} & -e^{-i\phi_2}\text{\ensuremath{\sin}}\frac{\theta_2}{2}\\
0 & e^{i\phi_2}\text{\ensuremath{\sin}}\frac{\theta_2}{2} & \text{\ensuremath{\cos}}\frac{\theta_2}{2}
\end{array}\right)$}\\
\hline  
\end{tabular}}
\caption{Experimental measurement settings for the 24 joint probabilities of 13 observables. We rotate the 3-dimentional coordinates to map at least two orthogonal observables on the axes of the measurement basis $\left\{\ket{1}, \ket{2}, \ket{3} \right\}$. Here $V_{i} ~(i=1,2, \cdots, 13)$ stands for the projection observables on the vectors $v_{i}$ shown in Fig. \ref{fig1:13}(a). These 16 settings contain all the joint measurements shown in Fig. \ref{fig1:13}(b) as edges. The rotations $R_1\left(\theta_1,\phi_1\right), R_2\left(\theta_2,\phi_2\right)$ are
realized by applying microwave pulses with the frequencies of $\omega_1$ and $\omega_2$, which are resonant to the transitions between $\ket{1}$ and $\ket{3}$ and between $\ket{2}$ and $\ket{3}$, respectively, where $\theta_1$, $\theta_2$ and $\phi_1$, $\phi_2$ are controlled by the duration and the phase of the applied microwaves. Here, $\alpha$ represents $0.392 \pi.$ Note that each observable will be transferred to the state $\ket{3}$ just before the measurement, since we only assign the values on the state $\ket{3}.$} 
\label{Tab1:Meassequence}
\end{table}

It is straightforward to measure the average values of single observables, $\avg{ A_{i}}$ in Eq. (\ref{eq1:OhYu}) and (\ref{eq3:OhYuSimple}). We rotate the measurement axis $v_{i}$ to the state $\ket{3}$ and then measure the probability of the state $P_{\ket{3}}$ ($=\avg{V_{i}}$), corresponding to the values of $\avg{ A_{i}}$, since $A_{i} = 1 - 2 V_{i}$. The probablity $P_{\ket{3}}$ is obtained by dividing the number of no photon events by the number of total repetitions. To measure the correlations, we first rotate two compatible vectors $v_{i}, v_{j}$ to the orthogonal states, for example, $\ket{1}$ and $\ket{2}$. Then we use the following modified equation to obtain the correlations $ \avg{ A_{i} A_{j}} = \avg{ \left(1-2 V_i\right) \left(1-2 V_j \right)}  = 1 - 2 \avg{V_i} - 2 \avg{V_j} + 4 \avg{V_i V_j}$, where $\avg{V_i}$ or $ \avg{V_j}$ is measured by the same method as that of a single observable, that is, transferring the population of state $\ket{1}$ or $\ket{2}$ to $\ket{3}$ and measuring the average probability of the state $\ket{3}$. 

The term $\avg{V_i V_j}$ is measured sequencially shown in Fig. \ref{fig2:Yb}(b). After mapping the axis of observables $V_{i}$ to $\ket{3}$ (no photon state), we would have two cases in our measurement: detecting certain number of photons or detecting no photons. If we observe photons, our state should not be $\ket{3}$, and we assign  zero $v(V_{i})=0$ on the observable $V_{i}$. If we observe no photons, our state should be $\ket{3}$ and we assign one $v(V_{i})=1$. When $v(V_{i})=0$, $v({V_i V_j})=0$ regardless of the result of $V_{j}$. Therefore, we stop the measurement. If $v(V_{i})=1$, we apply a swapping pulse between $V_{i}$ and $V_{j}$ and measure the state again. If we detect photons, which means $v(V_{j})=0$ and therefore, $v({V_i V_j})=0$. The correlation $V_{i} V_{j}$ has a value one only when we detect no photons at the second measurement.  We obtain the average of the correlation term $\avg{V_i V_j}$ by repeating the same experimental sequence. For the last terms in the equality (\ref{eq1:OhYu}), we apply the similar methods described above. First, the correlations $- \avg{ A_{i} A_{j} A_{k}}$ are expanded to $-1 + 2 \avg{V_{i}} + 2 \avg{V_{j}} + 2 \avg{V_{k}} - 4 \avg{V_{i}V_{j}} -4 \avg{V_{j}V_{k}} - 4 \avg{V_{k}V_{i}} + 8 \avg{V_{i}V_{j}V_{k}}$. We can ignore the terms $ \avg{V_{i}V_{j}V_{k}}$ because they should not have negative values and the inequality without the terms should be bounded by the same value, 25. Note that we do not throw away any measured data to construct the inequality, which ensures that the experiments are free from the detection loophole.  

\begin{figure}
\includegraphics[width=1\linewidth]{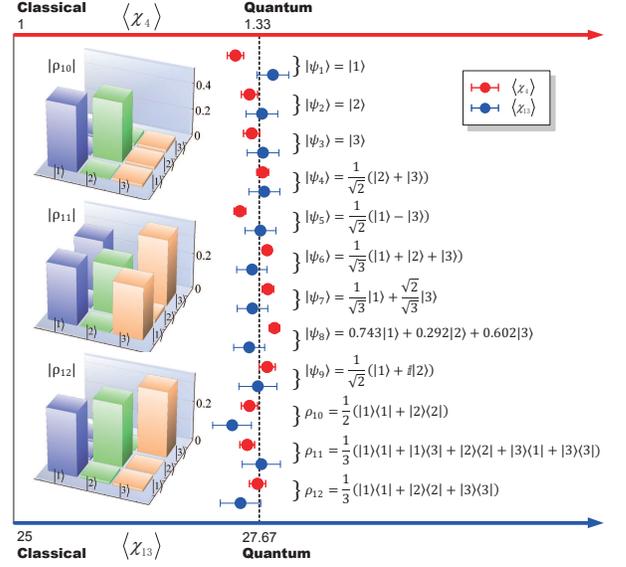}
\caption{ \label{fig3:Results} State-independent test of KS inequalities $\left\langle \chi_{4}\right\rangle$ and $\left\langle \chi_{13}\right\rangle$ for a qutrit system. The inequalities (\ref{eq1:OhYu}) and (\ref{eq3:OhYuSimple}) are examined with eleven different initial states including simple product states $\psi_{1}$ to $\psi_{3}$, superposition states at the axis of the observables $\psi_{4}$ to $\psi_{6}$, outside of the observables $\psi_{7}$ to $\psi_{9}$, and mixed states $\psi_{10}$ to $\psi_{12}$. We measure the density matrix of all states by performing qutrit state tomography \cite{Munro02} and confirm the prepared states with an average fidelity $99.5 \%$ for the pure states $\psi_{1}$ to $\psi_{9}$. The mixed state fidelities are around $97 \%$ shown in the figure. All raw measurements for $\avg{\chi_{13}}$ shown as blue filled circles violate the classical limit  25 by $15 \sigma$, confirming the quantum contextuality for various states. For each states, 480,000 realizations are used to obtain the $\avg{\chi_{13}}$. For the inequality (\ref{eq3:OhYuSimple}), the average value of all measured initial states is $\avg{\chi_{4}} = 1.328 (11)$.}
\end{figure}

The procedures of our experiments are as follows: after 1 ms Doppler cooling, the state of the ion is initialized to $\ket{3}$ by 3 $\mu$s standard optical pumping \cite{Olmschenk07}. The states are coherently manipulated by the microwaves $\omega_1$ and $\omega_2$ that are resonant to the transitions between $\ket{1}$ and $\ket{3}$, and between $\ket{2}$ and $\ket{3}$, respectively. The 2 $\pi$ times for both Rabi oscillations are adjusted to 29.5 $\mu$s, that is $\Omega_{1,2} = (2 \pi)~ 33.9$ kHz in frequency. The separation $\omega_2 - \omega_1$ = $(2 \pi)$ 7.6372 MHz with the magnetic field $B$= 5.455 G, which is large enough to ensure the independency of each Rabi oscillation.  

The 24 correlations of 13 observables of a given initial state are measured after rotating the observables on the one of qutrit states, shown in Table \ref{Tab1:Meassequence}. We prepare the measurement settings to map at least two orthogonal observables on the measurement basis simultaneously. The probabilities of $\ket{1}$ and $\ket{2}$ are obtained by mapping each state onto $\ket{3}$ and measuring the probability of $\ket{3}$. We repeat the same measurements 10,000 times for the same observables, which result in 480,000 repetitions for one initial state. After obtaining all the results of measurements, we apply the detection-error correction scheme with the maximally likelyhood method \cite{Chao12}. The state detection errors of no-photon state, $\ket{3}$ and of not detecting the $\ket{3}$ states are $1.0\%$ and $2.1\%$, respectively, with the discrimination threshold $n_{ph}=1$ shown in Fig. \ref{fig2:Yb}(b). 

We examine the state-independency of the inequalities (\ref{eq1:OhYu}) and (\ref{eq3:OhYuSimple}) by preparing eleven different initial states shown in Fig. \ref{fig3:Results} and repeating the measurements described in Table \ref{Tab1:Meassequence}. We verify that both inequalities are violated for simple input states $\psi_{1}$ to $\psi_{3}$, superposition states aligning to one of 13 observables $\psi_{4}$ to $\psi_{6}$, general superposition states not related to measurement axes $\psi_{7}$ to $\psi_{9}$, mixed states $\rho_{10}$ to $\rho_{12}$. The pulse sequences and state-tomography results are presented in the supplementary information. We observe the fidelities of the prepared states on average 99.5 $\%$ for the pure states. 

We clearly observe the violations of the inequaliteis (\ref{eq1:OhYu}) and (\ref{eq3:OhYuSimple}) for all the input states that we prepared, including mixed states as summarized in Fig. \ref{fig3:Results}. The average of $\avg{\chi_{13}} = 27.63 (17)$ and $\avg{\chi_{4}} = 1.328 (11)$, significantly larger than the limits of non-contextual models. For some of input states, $\avg{ \chi_{4}}$ and $\avg{\chi_{13}}$ are even larger than quantum bounds. However, the breaking of quantum bounds does not have serious meaning, since the results are within uncertainties of the measurements. We also observe the violations for all the states even without the correction of detection errors. However, the some of results are unphysical and violating even the quantum limit simply because of the detection errors. We provide all the detailed measurement results with and without detection error corrections in the supplementary information. 

In conclusion, our results show experimental measurements of an simplest three-level quantum system in an indivisible single atomic ion, which are incompatible to any non-contextual models. The results confirms the quantum contextuality for any states in three dimensions without entanglements or even superposition. Our experimental violations are not subject to the detection loophole. However, the sequencial measurements open the compatibility loophole \cite{Guhne10}. In our particular measurements, most of the observables in the inequalities are individual, which is irrelevant to the compatibilty loophole. And the correlations would be ignored if we require the same compatible algebric structure on the hidden variable models. Particularly, the violations of $\avg{\chi_{4}}$ conclude the quantum contextuality without detection and compatiblity loopholes.  

\subsection{Acknowledgement}
We thank Hyunchul Nha and Sewan Ji for the helpful discussion. We thank Emimly Lichko for carefully reading and improving the manuscript. This work was supported in part by the National Basic Research Program of China Grant 2011CBA00300, 2011CBA00301, 2011CBA00302, the National Natural Science Foundation of China Grant 61073174, 61033001, 61061130540. KK acknowledges the support from the Thousand Young Talents program. 

\bibliography{Contextuality}
\end{document}